# 400 kHz repetition rate THz-TDS with 24 mW of average power driven by a compact industrial Yb-laser

**C. MILLON[1*], S. HOUVER[2], C.J. SARACENO[1]**

[1]*Photonics and Ultrafast Laser Science, Ruhr University Bochum, 44801 Bochum, Germany*
[2] *Université Paris Cité, CNRS, Matériaux et Phénomènes Quantiques, F-75013 Paris, France*
**celia.millon@ruhr-uni-bochum.de*

**Abstract:** We demonstrate a high average power terahertz time-domain spectroscopy (THZ-TDS) set-up based on optical rectification in the tilted-pulse front geometry in lithium niobate at room temperature, driven by a commercial, industrial femtosecond-laser operating with flexible repetition rate between 40 kHz - 400 kHz. The driving laser provides a pulse energy of 41 µJ for all repetition rates, at a pulse duration of 310 fs, allowing us to explore repetition rate dependent effects in our TDS. At the maximum repetition rate of 400 kHz, up to 16.5 W of average power are available to drive our THz source, resulting in a maximum of 24 mW of THz average power with a conversion efficiency of ~ 0.15 % and electric field strength of several tens of kV/cm. At the other available lower repetition rates, we show that the pulse strength and bandwidth of our TDS is unchanged, showing that the THz generation is not affected by thermal effects in this average power region of several tens of watts. The resulting combination of high electric field strength with flexible and high repetition rate is very attractive for spectroscopy, in particular since the system is driven by an industrial, compact laser without the need for external compressors or other specialized pulse manipulation.

## 1. Introduction

Terahertz time domain spectroscopy has become a ubiquitous tool in many fields of science and technology [1–3]. This technique has immensely progressed in terms of THz energy per pulse, bandwidth and tunability. In particular, the increase in available THz pulse energy has opened the door to nonlinear spectroscopy directly in the THz range for a variety of scientific studies [4–6]. However, this progress has traditionally been achieved at the expense of the repetition rate of the source (<1 kHz), therefore the THz average power levels remain very low, typically in the few to hundreds of µW range. This is mostly due to limits of commonly used energetic Ti:Sapphire amplifiers as drivers, which have typically <<10 W of average power.

Higher THz average power is desired in a growing number of applications, for example to study water that has a large THz absorption [4] or in more difficult spectroscopic modalities such as multi-dimensional spectroscopy [7]. These, and other applications, would benefit from greater signal-to-noise ratios (SNR) and shorter measurement times, enabled by high repetition rates, while maintaining sufficient THz pulse energy, i.e. from sources with high average power. These experimental conditions were so far nearly exclusively available in accelerator facilities, at the expense very restrictive access and limited sensitivity due to pulse phase instabilities [8].

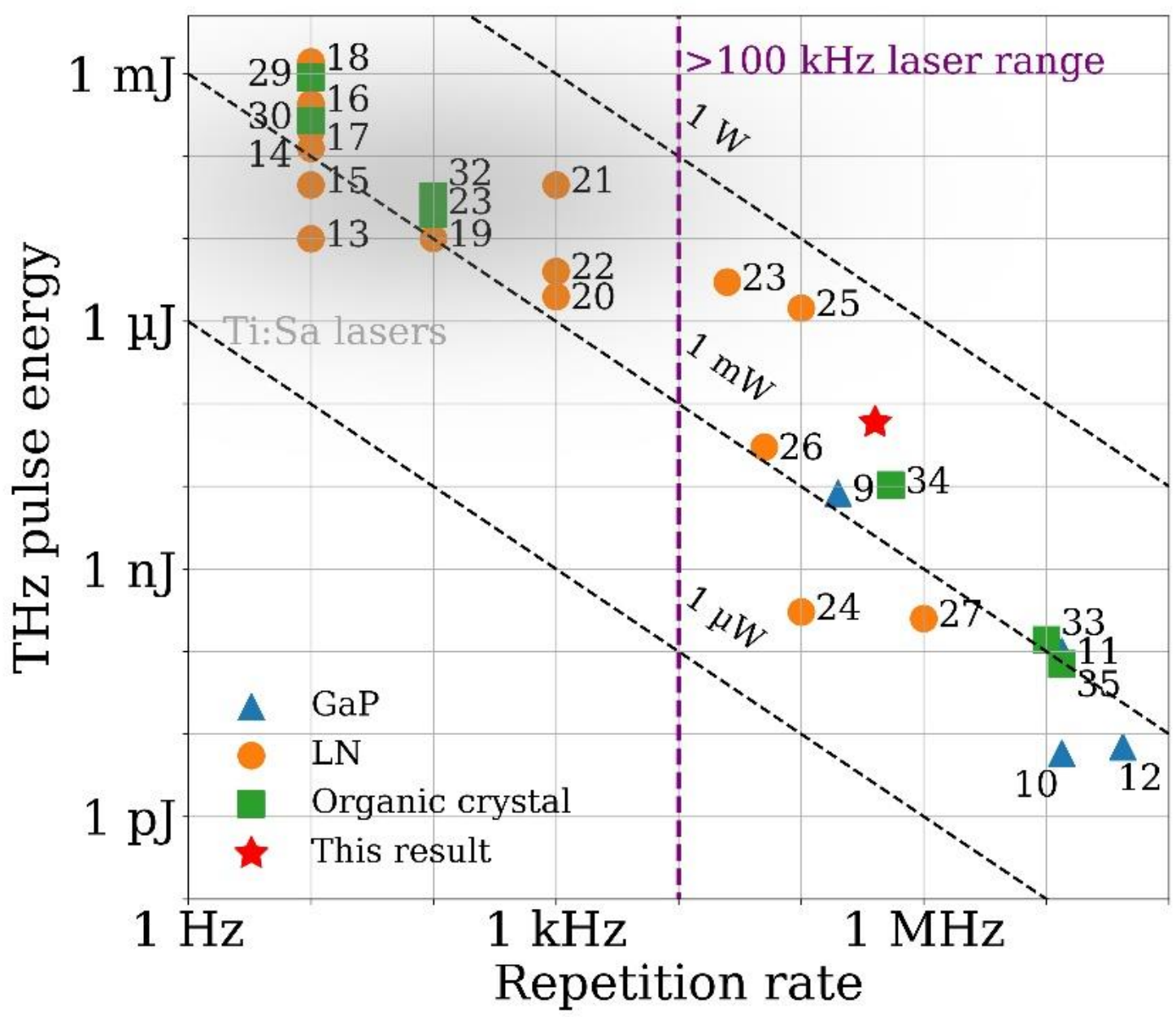

Figure 1 Overview graph of the THz pulse energy generated via optical rectification in different materials suitable for 1030 nm pumping: Lithium Niobate (LN), Gallium Phosphide (GaP) and organic crystals [9–35]. The diagonal dashed lines indicate the average power at 1 µW, 1 mW and 1 W. The vertical dashed line delimits the area of repetition rates below and above 100 kHz. The gray area displays the area targeted by Ti:Sapphire amplifiers.

One promising path to increase the average power of current THz sources, is to opt for ultrafast Yb-based laser systems (including oscillators and amplifiers) as driving sources, which have in the last decade surpassed the kilowatt average power level [36–39]. In particular, the repetition rate range between > 100 kHz and 1 MHz, i.e. between conventional Ti:Sapphire amplifiers and oscillators, provides an attractive combination of high repetition rate and high driving pulse energy. A large number of industrial and robust laser systems with these specifications, are commercially available. In spite of their potential, until very recently, the use of these lasers remained widely unexplored for THz generation via optical rectification (OR) in nonlinear crystals, partly because of the typically long pulse duration available from Yb-systems. However, recent progress in robust external high-power pulse compression techniques have enabled the emergence of high average power THz sources.

In this regard, the tilted-pulse front method in lithium niobate (TPF-LN) is very promising due to the high conversion efficiencies in the percent region, demonstrated at high driving pulse energy and low repetition rates [40]. First results have recently been reported showing high THz average power: using a 344 W slab amplifier Kramer et *al.* achieved 144 mW of THz power at 100 kHz repetition rate [25]; and aiming at much higher repetition rates, 66 mW of THz average power at 13.4 MHz were achieved using a 120 W modelocked thin-disk oscillator [41]. More recently, Guiramand et *al*. showed a 74 mW THz source at 25 kHz [23]. Collinear optical rectification has also been explored in the high-average power excitation regime, mostly to reach wider bandwidths albeit at typically lower conversion efficiency. Some relevant demonstrations include Mansourzadeh et *al.* [34] where 5.6 mW obtained at 540 kHz using organic crystal N-benzyl-2-methyl-4-nitroaniline (BNA) driven by a 10 W, 45 fs laser. Milliwatt level THz sources were also achieved at >10 MHz repetition rates using organic crystals [33,34]. With Gallium Phosphide crystal (GaP), Meyer et *al.* reached 1.35 mW of THz average power at 13.4 MHz [11] and very recently, Nilforoushan et *al.* demonstrated a broadband 1.66 mW THz source at 200 kHz [9]. Most of the studies up to now where focused on rather moderate repetition rate few tens of kHz and rather high energy per pulse in the range of the mJ. We are studying optical rectification in LN for low energy per pulse and high repetition rate. The low energy necessitates small pump beam sizes in order to achieve a sufficiently high peak intensity for optical rectification. In this paper the pump hitting the LNO crystal is in the order of the millimeter. This combination of small beam sizes and high average power constitutes a new excitation regime: as the average density increases, thermal effects are likely to happen. Simulation of such thermal effects resulting from this excitation regime are complex because multi-factors, as Wulf pointing out [42]. Thus, a direct experimental study such as the present one is of interest for future works in this regime."

An overview of all these results is shown on Fig.1 where we display the energy per THz pulse as a function of the repetition rate, thus the average power is illustrated by diagonal lines. Please note that this overview disregards the THz bandwidth and peak electric field of the different sources, which can significantly affect the choice of a certain technique for a given application. While for nonlinear spectroscopy it is clear that the peak electric field is of critical importance, we argue here that pulse energy is a fairer metric to compare different literature results, given the well-known difficulties in precisely calculating peak electric field and exact generated bandwidth, which we discuss later in this paper.

In this context, we focus our attention on the TPF-LN method to reach high conversion efficiency in the promising repetition rate region (100 kHz-1 MHz), using compact and industrially mature Yb-lasers. We demonstrate a THz-source based on the TPF-LN at room temperature, driven by 16.5 W of average power from a commercial, compact fiber-laser delivering 310 fs pulses at 400 kHz repetition rate. Under optimized conditions, we obtain 24 mW of THz average power with a conversion efficiency of ~ 0.15 %, closely following the predictions of Wulf et *al.* [42] about TPF-LN operated at moderate pulse energies. We carefully characterize the electric field achieved in our setup and discuss the variations measured using different methods. We highlight that this system is very compact, requires no external compressors and is directly driven by a turn-key industrial laser. Moreover, this flexible laser system allows us to changes the repetition rate and thus explore its influence (40 kHz to 400 kHz) on the performance of our TDS. Interestingly, our results show that in this excitation regime, thermal effects in LN do not affect the generated THz electric field. Our system is therefore very attractive for applications in spectroscopy where one might want to have repetition rate flexibility with a high SNR. As an example, Novelli et *al.*[43] carried an non-linear spectroscopy experiment on water samples. They have observed an optical, an acoustic and a thermal response at different time scale due to in particular to the repetition rate of the pump. Thus, the difference in time scale and physical phenomenon requires flexibility in repetition rate. The system is also of interest for nonlinear spectroscopy requiring a few tens of kV/cm [44] .

## 2. Experimental set-up

An overview of the experimental set-up is given in Fig.2. Our driving laser is a femtosecond industrial laser typically used in laser material processing applications, (TRUMPF Trumicro 2000) which delivers 310 fs pulses at 1030 nm with up to 19 W of average power at 400 kHz repetition rate. The footprint of the laser is 62.5 cm by 37.5 cm and a picture is shown in the inset of our experimental setup. After the transmission grating, the pump power measured is 16.5 W (power impingent on the crystal), resulting in ~ 41 µJ of energy per pulse available for the optical rectification process. The laser repetition rate can be varied at constant pulse energy between 400 kHz and 40 kHz. The pulse duration and the spot size remain the same regardless of the variation of repetition rate. Thus, the average power hitting the LN crystal change correspondingly from 1.65 W to 16.5 W, allowing us to explore potential repetition rate dependent effects.

The pulses are divided into the probe and pump arms using a 1% output coupler (OC). To adjust the laser power and pulse energy applied on the crystal, a combination of a thin-film polarizer (TFP) and a half-waveplate (λ/2) is used. The THz waves are generated through optical rectification in a 0.6% MgO-doped stoichiometric LN trapezoid following the guidelines presented in [42]. The spot size of the pump inside the crystal, optimized for the highest conversion efficiency, is approximately 1.3 mm ($1/e^2$ diameter). A pulse front tilt of 63° is used to achieve velocity matching between pump and THz fields. This is realized using a transmission grating (transmission of 90 % at the first diffraction order) with 1600 lines/mm set at an angle of incidence of 55.5° in combination with a 2” aspherical lens, which images the grating into the crystal with a focal length of 87 mm. The THz beam exits the front surface of the trapezoid; it is collected and refocused by two 2” off-axis parabolic mirrors of 50 mm focal length each. The pump beam is reflected off the front surface of the LN trapezoid by total internal reflection. Both the beam trap and the crystal mount are water cooled. Throughout the THz generation path, a chopper wheel at 980 Hz in the pump arm is implemented for lock-in detection. A fast delay line (ScanDelay 15 ps, APE GmbH), oscillating at 0.4 Hz, is used to display the photodiode signal to the lock-in amplifier. We characterize the THz radiation either by evaluating the energy per pulse or by plotting the THz electric field and the corresponding spectrum. In this paper, the energy per pulse is calculated by dividing the power measured by a pyroelectric power meter (Gentec, THZ12D-3S-VP-D0) by the repetition rate. We note that the average power values measured were reliably compared to a THz power meter calibrated in this spectral region by the PTB (THz 20, SLT GmbH). The THz electric field trace is obtained by an electro-optical-sampling (EOS) set-up, using a 0.5 mm GaP crystal, followed by a quarter-wave plate, a Wollaston prism, and a balanced photodiode (PD). Thin layers of teflon (~200 µm) are used in the collimated part of the THz to filter the residual near infrared radiation. The path of the THz beam is purged with dry nitrogen to approximately 10 % relative humidity in order to reduce water vapor absorption in the air.

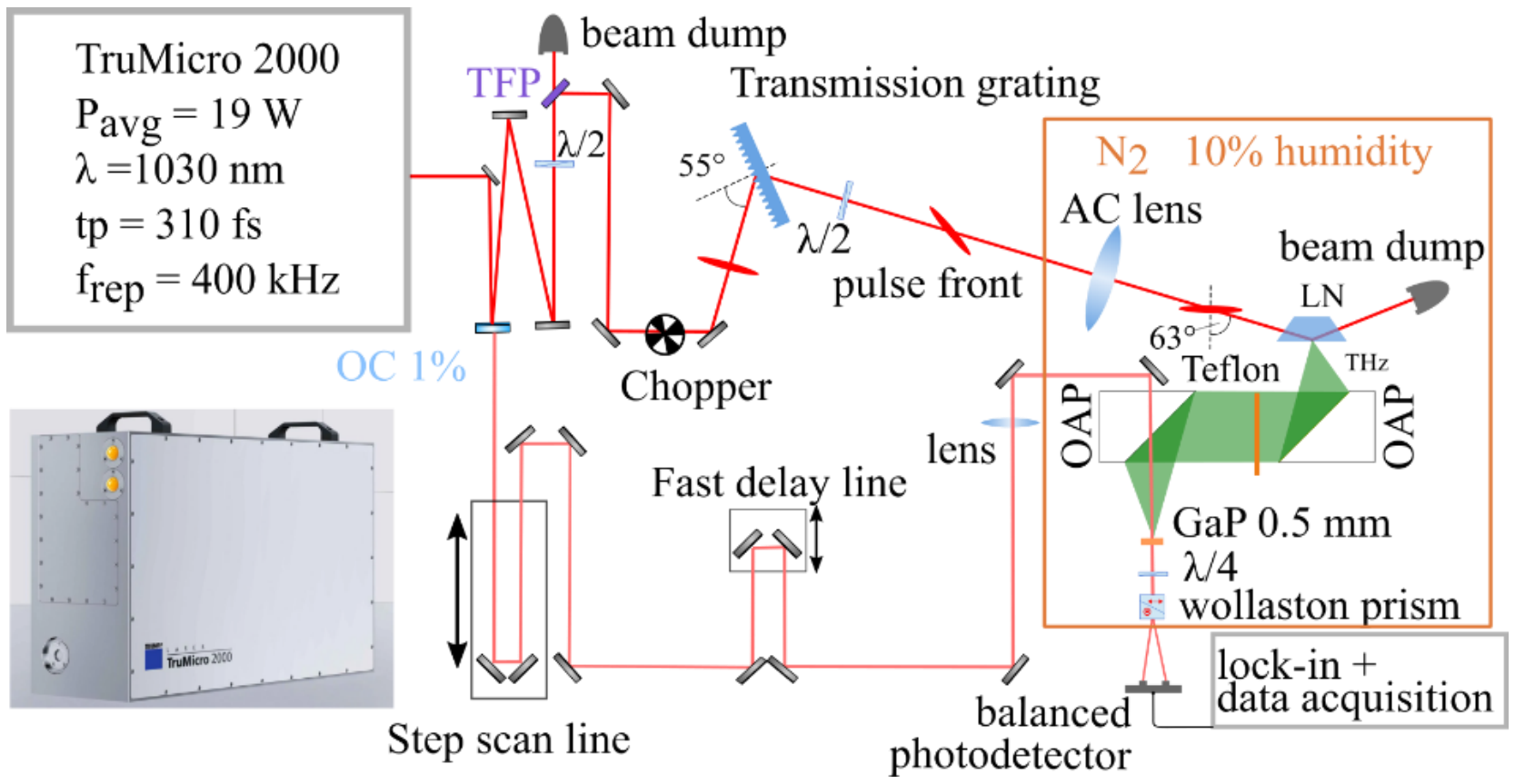

Figure 2 THz-TDS setup THz generation in lithium niobate based on tilted pulse front technique. TFP: Thin film polarizer, OC: Output coupler, OAP: Off Axis Parabolic mirror, AC: Achromatic lens, GaP: Gallium Phosphide, LN: Lithium Niobate. The inset shows a schematic of Trumicron 2000.

## 3. Results and Discussion

### Repetition rate influence on THz generation

Fig. 3.a displays THz transients acquired by EOS in the time domain for 400 kHz and 40 kHz repetition rate under purged conditions, for the same single pump pulse energy of 41 µJ. The THz electric field remains the same for the lowest and highest repetition rates. Fig. 3.b shows the corresponding normalized power spectra on a logarithmic scale. The spectra are centered in the vicinity of ~0.8 THz. We average over 60 traces within only 76 s, resulting in a dynamic range (DR, i.e. ratio of the maximum amplitude and the root mean square of the noise floor) of ~58 dB for 40 kHz repetition rate and ~68 dB for 400 kHz. The DR increases by 10 dB for an increase in repetition rate by a factor of 10 showing the advantage of higher repetition rates to improve this DR. We underline that for experiments where acquisition time rather than peak dynamic range is critical, this setup would allow to operate at 10 times faster at 400 kHz without any change in performance (DR, bandwidth, and peak electric field).

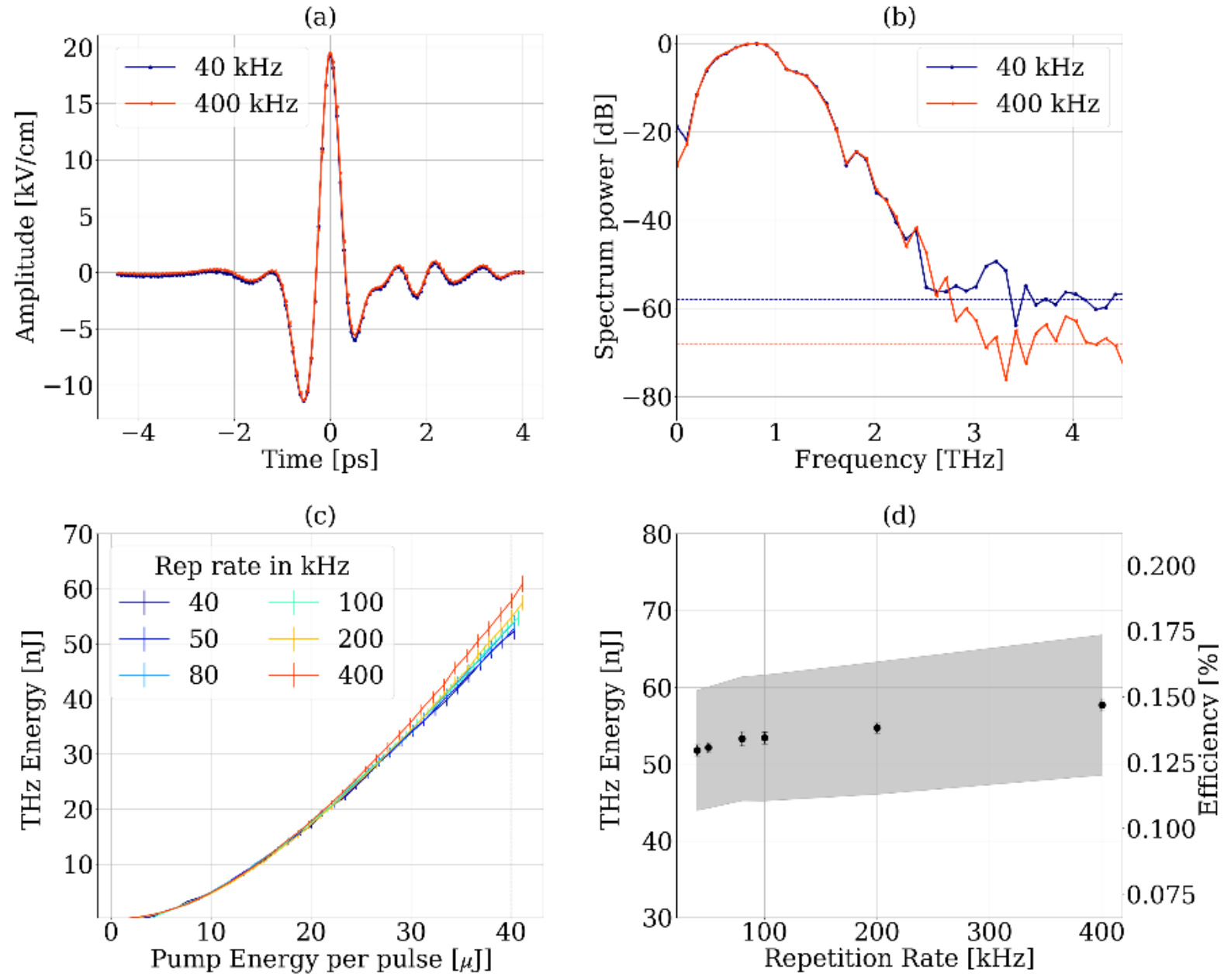

Figure 3 (a) THz electric field temporal trace for two repetition rates at 40 and 400 kHz, with the same pump energy of 40 µJ. (b) Corresponding THz spectra. (c) Variation of the THz energy for different repetition rate. (d) THz energy at a constant pump energy of 40 µJ against repetition rate. The corresponding efficiency is also given on the second y-axis.

Fig. 3.c depicts the THz pulse energy against the pump pulse energy for different repetition rates, evaluated with the powermeter. The highest THz average power is obtained for 41 µJ of pump pulse energy at the highest repetition rate of 400 kHz. Considering the black textile on the power meter and the teflon layers transmitting 67 % and 93 %, respectively, the measured power is 24 mW (± 15 % according to the power meter calibration), corresponding to a conversion efficiency, i.e. ratio of the THz power over pump power, of ~ 0.15 %. Compared to recent results obtained in this repetition rate region we note that our setup reaches significantly higher efficiency than the setup of Kramer et *al*. [25] which was most likely in presence of strong thermal effects and possibly in non-optimized pulse duration regime. This power result is also more than an order of magnitude higher than recent collinear results in GaP [45] (1.66mW) albeit at more moderate bandwidth spanning up to 2.5 THz. From fig.3.c, one can observe that energy curves are superimposed for the whole range of repetition rates within the measurement accuracy. Fig. 3.d shows the THz energy, as well as the conversion efficiency, for different repetition rates, at a constant pump energy of 40 µJ. The error bars correspond to the standard deviation, the shaded area indicates the linear response of the THz power meter, which can be up to 15 % off. Within the uncertainty of the power meter, the generated THz energy remains constant with varying repetition rates from 40 to 400 kHz. This is in good agreement with the unchanged THz transients and their spectra with the repetition rate as shown in fig.3.a and 3.b. Based on these two independent measurements, EOS and power meter, we conclude

that no thermal load is affecting the THz generation within the LN crystal. These laser conditions with an energy per pulse significantly inferior to the millijoule level and a repetition rate several orders of magnitude higher than the kHz, are therefore very appealing for improving the SNR of the THz signal in a THz-TDS set-up, and also for setups where a variation of the repetition rate is desirable for example to study cumulative effects.

## Characterization of THz electric field

For nonlinear spectroscopy, the critical parameter is the achievable electric field. Its estimation is often done by either estimating the measured modulation of the photodiode using the EOS detection, or by assuming a gaussian shape of the THz beam and evaluating the area of the beam with the THz camera in combination with a power measurement. These two evaluations are subjected to a lot of critical parameters: alignment, size of the probe beam, spatial overlap between the THz and the probe beams, variations of surface quality of the EOS crystal [46]. The evaluation with the THz camera and the power meter is subjected to the power meter uncertainty, the uncertainty in the evaluation of the beam area of a broadband beam, and the estimation of the pulse duration. Moreover, as the image of the THz beam is taken at the focus of the second parabolic mirror, the beam shape is also subjected to the aberrations of the imaging system made by the two parabolic mirrors. The evaluation with the photodiode implies a perfect alignment of the EOS set-up to give a proper estimation of the electric field. Thus, this method most likely gives an underestimation of the electric field. Nevertheless, it is valuable to have an estimation given by these two methods. In order to provide these estimations and compare our work to the existing literature, we evaluate the field using these two methods and present them below. However, we note that on the one hand the focusing conditions could be optimized with careful design of specialized optical elements, which affects the peak electric field; and on the other hand, we believe that ultimately the nonlinear experiment performed with a well-studied sample would give a cross-checked information on the electric field strength.

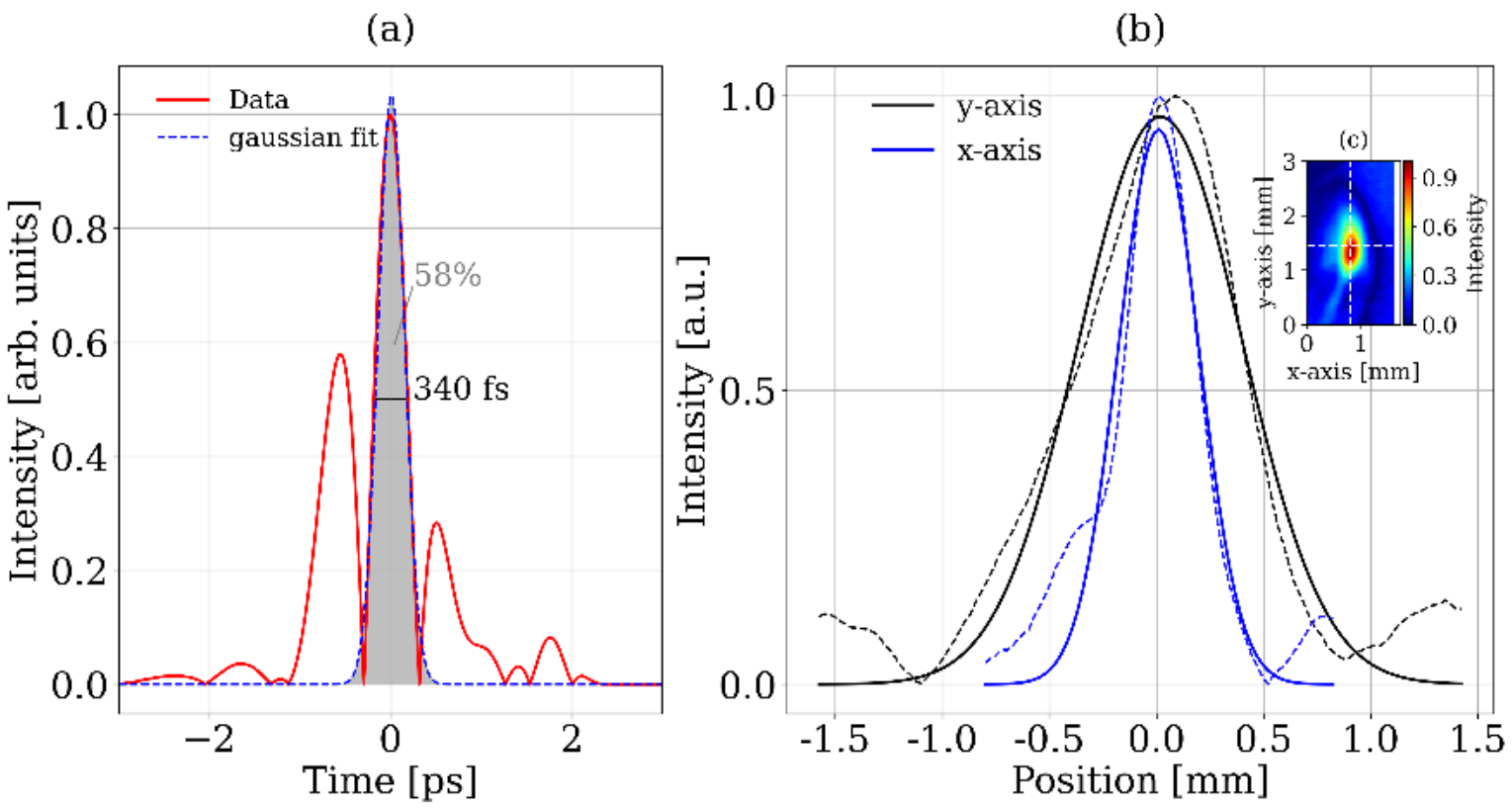

Figure 4 (a) Intensity profile of the THz pulse. The main cycle of the electric field carries 58% of the pulse energy and is used to estimate the FWHM. (b) Lineouts of the THz beam along the axis indicated in (c). The dashed lines represent the raw data. (c) THz beam at the focus of the second parabolic mirror acquired with Rigi Camera -Swiss THz.

The first technique we use is based on measuring the modulation of the photodiode current due to the electro-optic response of the nonlinear crystal. The THz electric field is given by the following equation [47]:

$$E_{THz} = \frac{\Delta I}{I} \frac{\lambda}{2\pi \cdot r_{41} \cdot n^3 \cdot L \cdot t_{fresnel} \cdot t_{teflon}} \quad (1)$$

Where $\lambda$ and $n$ are respectively the wavelength and the refractive index of the pump, $t_{fresnel}$ is the transmission coefficient in amplitude of the THz beam at the air-GaP interface and is equal to 0.49 in this case, $t_{teflon}$ is the transmission coefficient of the teflon, $L$ is the thickness of the GaP crystal. $\Delta I$ is the current detected by both balanced photodiodes when the THz field is present. $I$ the total amount of the current hitting the photodiodes when the THz is off. The electro-optic coefficient $r_{41}$ of GaP at 1030 nm equals to 0.47 pm/V [48], using the equation (1), a peak field of 19.5 kV/cm is estimated. By comparison, the electric field can also be evaluated by measuring the THz spot size on the crystal with a THz camera. To characterize the THz pulse, the Gaussian envelope of the main peak of the THz time trace shown in Fig. 4.a is used. The intensity of the main peak can be calculated as follows:

$$I_{THz} = 0.94 \frac{W_{THz}}{A_{eff} \tau_{THz}} \tag{2}$$

Where, $A_{eff}$ is the effective area calculated using the diameter at $1/e^2$ level in the focal plane of the second parabolic mirror and $\tau_{THz}$ is the THz pulse duration at full width half maximum (FWHM). Based on the work of Sitnikov et *al.* [49], the pulse duration is evaluated by fitting a gaussian on the half cycle carrying 58% of the intensity as shown on fig.4a. The calculated half period temporal width is about 340 fs. Fig. The peak intensity ($I_{THz}$ in the equation (2)) is estimated for this main half-period (see dashed gaussian fit in Fig. 4.a). Moreover, $W_{THz}$ indicates the effective energy of this main peak and correspond to the power measured by the pyroelectric detector and divided by the repetition rate, 400 kHz, giving 60 nJ. Fig. 4.b shows lineouts along the x- (horizontal, plane of tilt pulse front) and y-direction (vertical). The resulting $1/e^2$ diameters are 1.08 mm and 0.53 mm; the area of the beam is estimated to 1.7 $mm^2$. The electric field is then given by the following formula:

$$E_{THz} = \sqrt{2Z_0 I_{THz}} \tag{3}$$

Where $Z_0$ is the impedance of vacuum. The obtained value of the THz electric field using equation (3) is 67 kV/cm.

The two estimations are within the same order of magnitude and the factor 3 between the two evaluations are commonly seen in the literature using these two methods [23]. It is to be noticed that the focusing conditions could be further optimized for spectroscopy applications, using for example 100 mm and 50 mm focal length parabolic mirrors, possibly achieving significantly higher fields. The THz fields reached are sufficient for many nonlinear experiments, with the additional benefit of a high dynamic range exceeding 70 dB and short measurement times. In addition to this, the source is repetition rate tunable, making this set-up a unique, very flexible tool for THz spectroscopy.

## 4. Conclusion

In conclusion, we demonstrate THz generation by optical rectification in 0.6% MgO-doped stoichiometric lithium niobate trapezoid pumped by a commercial, high- and flexible-repetition rate Yb-laser, and present a detailed study of THz energy against the available repetition rates. In optimized conditions at 400 kHz repetition rate, we reach a maximum THz power of 24 mW with a spectrum extending to 2.5 THz and a spectral dynamic range of ~70 dB. To the best of our knowledge, this is the highest THz average power obtained with optical rectification for this range of repetition rate and energy per pulse (hundreds of kHz and tens of µJ). Furthermore, we show that thermal effects are not a limiting factor in this laser category (tens of µJ pulses and hundreds of kHz), and thus that our TDS can be operated between 40 – 400 kHz without modifying field strength or bandwidth. This source is very attractive for THz-TDS experiments where high power, high repetition rate and high dynamic range is advantageous. We believe it can find widespread applications in spectroscopy, in particular since it is driven by a compact commercial femtosecond laser system without any specialized external pulse manipulation needed. In the near future, we plan to optimize the THz beam focusing conditions by using different parabolic mirrors. We believe this can lead this source into the hundreds of kV/cm field regime, necessary for strongly nonlinear applications.

### Acknowledgements

This project received funding from the European Union's Horizon 2020 research and innovation program under the Marie Skłodowska-Curie grant agreement No 801459 - FP-RESOMUS and was funded by the Deutsche Forschungsgemeinschaft (DFG) under Germany's Excellence Strategy - EXC 2033 - 390677874 - RESOLV. We acknowledge the DFG Open Access Publication Funds of the Ruhr-Universität Bochum.

### Disclosures

The authors declare no conflict of interest.

### Data availability

The data are available from the authors upon reasonable request.